\documentclass[preprint,preprintnumbers,showpacs,aps,amssymb]{revtex4}

\usepackage{graphicx}
\usepackage{bm}
\usepackage{amsmath}

\def\calL{{\cal L}}
\def\calO{{\cal O}}

\def\cbar{{\bar c}}
\def\psibar{{\bar\psi}}
\def\nubar{{\bar\nu}}
\def\ebar{{\bar e}}

\def\phat{{\hat p}}
\def\Phat{{\hat P}}
\def\shat{{\hat s}}

\def\nn{\nonumber}

\begin{document}
\title{More than one ultimate speed and superluminal neutrinos}
\author{Jong-Phil Lee}
\email{jplee@kias.re.kr}
\affiliation{Institute of Convergence Fundamental Studies, Seoul National University of Science and Technology, Seoul 139-931, Korea}
\affiliation{Division of Quantum Phases $\&$ Devices, School of Physics, Konkuk University, Seoul 143-701, Korea}

\begin{abstract}
It is suggested that recent superluminal neutrinos from the OPERA collaboration might indicate
that there are other ultimate speeds than usual speed of light in our universe.
The leptonic sector of the standard model (SM) is reformulated incorporating with new ultimate speeds.
In minimal cases where there is another maximum speed $c'$ which equals to the speed of the OPERA neutrinos,
new effects would appear at the level of $\calO(10^{-5})$.
Improved precisions for the electroweak observables would check the validity of this scenario.
\end{abstract}

\maketitle
Recently the OPERA collaboration has reported that neutrinos traveling faster than light are observed.
The result is so striking that a surge of discussions has arisen both inside and outside the 
science community, and the collaboration reanalyzed the result to update as \cite{OPERA2}
\begin{equation}
\frac{v_\nu-c}{c}=(2.37\pm0.32^{+0.34}_{-0.24})\times10^{-5}~.
\label{OPERA}
\end{equation}
After the announcement there have been a lot of works to explain or refute the results.
Interpretations in favor of the result include the Lorentz violation \cite{LIV}, 
medium effects \cite{medium}, and extra dimensions \cite{XD}, to name a few.
\par
In this paper we provide a new explanation for the superluminal neutrinos.
We assume that while electrons and photons have the usual maximum speed $c$, 
there exist other ultimate speeds for neutrinos (and possibly also for other particles) 
and analyze what kinds of effects are expected.
A similar idea was given in \cite{Schreck}.
For example, one may think that only neutrinos have another ultimate speed $c'>c$ 
and all other particles do not exceed the usual speed of light $c$.
It will be shown that in this case the effect of $c'$ would appear in the Fermi constant.
In a more general case other particles also have $c'$ as an ultimate speed and affect well-known
physical quantities such as weak mixing angle.
Assuming that $c'=v_\mu$, the new effects would appear at order of $\calO(10^{-5})$ from Eq.\ (\ref{OPERA}).
But many electroweak observables up to now have larger uncertainties,
so the "$c'$-scenario" might be a good explanation for the OPERA result.
\par
In incorporating $c'$ in the Lagrangian, it is not convenient to use 
$\partial_\mu=(\partial_t/c,~{\vec\nabla})$ since $c$ is contained within it.
One must use other $\partial_\mu$'s for particles with different $c$'s.
Instead it is convenient to use the momentum operator $\phat_\mu$ in place of 
$i\hbar\partial_\mu\to\phat_\mu$.
\par
We start with the observation that the Lagrangian for a massless Dirac fermion can be written as
(explicitly keeping $c$ and $\hbar$)
\begin{equation}
i\hbar c\psibar\gamma^\mu\partial_\mu\psi
=(\sqrt{c}\psibar)\gamma^\mu\phat_\mu(\sqrt{c}\psi)~,
\end{equation}
where we use the momentum operator $\phat_\mu$ in place of $i\hbar\partial_\mu$.
Now we assume that neutrinos have different ultimate speed, $c'$, instead of the speed of light $c$.
The Lagrangian for a neutrino and an electron is 
\begin{eqnarray}
\calL_f&=&\nubar_L\gamma^\mu\phat_\mu c'\nu_L+\ebar_L\gamma^\mu\phat_\mu c e_L\nn\\
&=&\psibar_L\gamma^\mu\phat_\mu\psi_L~,
\end{eqnarray}
where $L$ denotes left-handed and $\psi_L^T=(\sqrt{c'}\nu_L~\sqrt{c}e_L)$.
The covariant form of $\calL_f$ can be easily obtained by replacing
\begin{equation}
\phat_\mu\to\Phat_\mu=\phat_\mu+\frac{gA_\mu^a}{c_a}\tau^a~.
\end{equation}
In this expression we allow different "speed of light" $c_a$'s in general for $A_\mu^a$.
For the $SU(2)_L\times U(1)_Y$ gauge theory,
\begin{equation}
\Phat_\mu=\phat_\mu+\frac{g_2A_\mu^a}{c_a}\tau^a+\frac{g_1B_\mu}{c_B}Y~.
\end{equation}
Gauge bosons become massive through the usual Higgs mechanism.
When the scalar field $\phi$ gets its vacuum expectation value
$\langle\phi^T\rangle=(0~v/\sqrt{2})$,
the mass terms appear as
\begin{equation}
|\Phat_\mu\phi/\hbar|^2
=\frac{1}{2\hbar^2}\frac{v^2}{4}\left[
g_2^2\frac{c_1^2+c_2^2}{c_1^2c_2^2}W_\mu^+W^{-\mu}+
\left(\frac{g_1^2}{c_B^2}+\frac{g_2^2}{c_3^2}\right)(Z_\mu^0)^2\right]
+({\rm other~terms})~,
\label{massterm}
\end{equation}
where
\begin{eqnarray}
W_\mu^\pm&=&\frac{c_1c_2}{\sqrt{c_1^2+c_2^2}}\left(
 \frac{A_\mu^1}{c_1}\mp i\frac{A_\mu^2}{c_2}\right)~,\\
Z_\mu^0&=&\frac{1}{\sqrt{g_1^2/c_B^2+g_2^2/c_3^2}}\left(
 \frac{g_2}{c_3}A_\mu^3-\frac{g_1}{c_B}B_\mu\right)~.
\end{eqnarray}
From Eq.\ (\ref{massterm}) one can extract the $W$ and $Z$ boson masses
\begin{eqnarray}
m_Wc_W&=&g_2\frac{v}{2}\sqrt{\frac{c_1^2+c_2^2}{2c_1^2c_2^2}}~,\\
m_Zc_Z&=&\frac{v}{2}\sqrt{\frac{g_1^2}{c_B^2}+\frac{g_2^2}{c_3^2}}~,
\end{eqnarray}
where $c_{W,Z}$ are the ultimate speeds associated with the $W, Z$ bosons.
\par
The fermionic Lagrangian becomes
\begin{eqnarray}
\calL_f
&=&\nubar_L\gamma^\mu\phat_\mu c'\nu_L+\ebar_L\gamma^\mu\phat_\mu c e_L\nn\\
&&+g_2\frac{\sqrt{c_1^2+c_2^2}}{2c_1c_2}\left(
 \sqrt{c'}\sqrt{c}\nubar_L\gamma^\mu W_\mu^+e_L+\sqrt{c'}\sqrt{c}\ebar_L\gamma^\mu W_\mu^-\nu_L
 \right)\nn\\
&&+\frac{g_2}{c_3\cos\theta'_W}\left[
 \frac{1}{2}c'\nubar_L\gamma^\mu Z_\mu^0\nu_L
 +c\ebar_L\gamma^\mu Z_\mu^0\left(-\frac{1}{2}+\sin^2\theta'_W\right)e_L\right]\nn\\
&&+ce'\ebar_L\gamma^\mu A_\mu(-1)e_L~,
\label{Lf}
\end{eqnarray}
where
\begin{equation}
A_\mu
=\frac{1}{\sqrt{g_1^2/c_B^2+g_2^2/c_3^2}}\left(\frac{g_1}{c_B}A_\mu^3
 +\frac{g_2}{c_3}B_\mu\right)~.
\end{equation}
Here $\theta'_W$ is the mixing angle between $Z_\mu^0$ and $A_\mu$ as in the usual SM,
\begin{equation}
\cos\theta'_W=\frac{g_2/c_3}{\sqrt{g_1^2/c_B^2+g_2^2/c_3^2}}~,~~~
\sin\theta'_W=\frac{g_1/c_B}{\sqrt{g_1^2/c_B^2+g_2^2/c_3^2}}~,
\end{equation}
and
\begin{equation}
e'=\frac{1}{\sqrt{g_1^2/c_B^2+g_2^2/c_3^2}}\frac{g_1}{c_B}\frac{g_2}{c_3}~.
\end{equation}
\par
Usually the Fermi constant is extracted from the muon decay.
From the charged current of $\calL_f$ in Eq.\ (\ref{Lf}), one can calculate the muon lifetime
(assuming that muon's maximum speed is $c$)
\begin{equation}
\tau=
\left(\frac{192\pi^3\hbar^7}{m_\mu^5c^4}\right)
\frac{32}{c^2\hbar^6}
\left(\frac{m_Wc_W}{g_2}\right)^4\left[\frac{2c_1^2c_2^2}{(c_1^2+c_2^2)c'c}\right]^2~,
\end{equation}
so the Fermi constant is given by
\begin{equation}
G_F'=\frac{\sqrt{2}}{8}(c\hbar^3)\frac{g_2^2}{m_W^2c_W^2}
\left(\frac{c_1^2+c_2^2}{2c_1^2c_2^2}cc'\right)~.
\label{GF}
\end{equation}
Note that all the SM results are restored when all kinds of $c$'s are equal to the 
usual speed of light, $c$.
\par
To be more specific, consider a simple case of 
$c'=(1+\delta)\ne c_W=c_Z=c_1=c_2=c_3=c_B=c=1$ (case 1).
In case 1, $\theta_W'=\theta_W$, and $e'=e$.
We assume that $c'=1+\delta~(\delta\ll 1)$ is the speed of the OPERA neutrinos.
In this case the effect of $\delta$ appears in $G_F$ as
\begin{eqnarray}
G_F'&=&\frac{\sqrt{2}}{8}\frac{g_2^2}{m_W^2}(1+\delta)\nn\\
&=&\frac{\sqrt{2}}{8}\frac{e^2}{m_W^2\sin^2\theta_W}(1+\delta)
~.\label{GF1}
\end{eqnarray}
Note that the precision required to see the effect of $\delta$ is $\calO(10^{-5})$.
The most precise measurement of the $W$ mass is given by recent CDF \cite{CDF2} 
and D0 \cite{D0} collaborations,
\begin{eqnarray}
m_W&=&80.387\pm 0.019~{\rm GeV~(CDF)}~,\\
&=&80.367\pm 0.026~{\rm GeV~(D0)}~,
\end{eqnarray}
thus the errors are about one order of magnitude larger than $\delta$.
\par
On the other hand the measured value of $m_Z$ and $G_F'$ is \cite{PDG}
\begin{eqnarray}
m_Z&=&91.1876\pm 0.0021~{\rm GeV}~,\label{mZ}\nn\\
G_F'&=&(1.16637\pm 0.00001)\times 10^{-5}~{\rm GeV}^{-2}~,\label{GF}
\end{eqnarray}
so their uncertainties are comparable to or smaller than $\delta$.
Since $m_W/m_Z=\cos\theta_W$ in case 1, one can determine $\sin\theta_W$ from $G_F'$ as
\begin{equation}
(1-\sin^2\theta_W)\sin^2\theta_W=\frac{\pi\alpha}{\sqrt{2}G_F'm_Z^2}(1+\delta)~.
\label{s2W1}
\end{equation}
In the SM the value of $\sin^2\theta_W$ at the scale of $m_Z$ determined in this way 
(i.e., when $\delta=0$) is
$\sin^2\theta_W=0.23108\mp0.00005$ \cite{PDG}.
For $\delta\ne 0$, we have from Eq.\ (\ref{s2W1}) $\sin^2\theta_W=0.23109\mp 0.00005$.
Note that the difference between the two values is within the error.
The renormalized quantity of $\sin^2\theta_W$ at $m_Z$ with minimal subtraction, 
$\sin^2{\hat\theta}_W\equiv\shat^2_Z$,
is obtained by 
$\shat^2_Z=\cbar\sin^2\theta_W(m_Z)$
where $\cbar=1.0010\mp0.0002$ \cite{PDG}.
In case 1, one has $\shat^2_Z=0.23132\mp 0.00007$.
The value can be compared with the Particle Data Group result 
$\shat^2_Z=0.23116\pm00013$ \cite{PDG}.
For a complete comparison one must implement the whole procedure to fix $\shat^2_Z$ 
in the "$c'$-scenario" but it is beyond the scope of this version of the paper.
\par
Another interesting case is 
$c_W=c_Z=c_1=c_2=c'=(1+\delta)\ne c_3=c_B=c=1$ (case 2).
In case 2, $\theta'_W=\theta_W$, $e'=e$, and other quantities are
\begin{eqnarray}
m_W&\simeq&g_2\frac{v}{2}(1-2\delta)~,\label{mW2}\\
m_Z&\simeq&\frac{v}{2}\sqrt{g_1^2+g_2^2}(1-\delta)~,\label{mZ2}\\
G_F'&\simeq&\frac{\sqrt{2}e'^2}{8m_W^2\sin^2\theta_W}(1-3\delta)~.\label{GF2}
\end{eqnarray}
Note that the electroweak observables have very different expressions in case 2 compared with case 1.
Combining Eqs.\ (\ref{mW2}) and (\ref{mZ2}), a remarkable result comes out:
\begin{equation}
\frac{m_W}{m_Z}\simeq (1-\delta)\cos\theta_W~.
\label{cW2}
\end{equation}
From the measured values of $m_W$, $m_Z$, and $\delta$, one can extract $\theta_W$.
The coupling constants $g_1$ and $g_2$ are fixed through 
$g_1=e/\cos\theta_W$ and $g_2=e/\sin\theta_W$.
Or as before the value of $\sin^2\theta_W$ is extracted from the measurement of the Fermi constant
via Eqs.\ (\ref{GF2}) and (\ref{cW2}).
In this case
\begin{equation}
(1-\sin^2\theta_W)\sin^2\theta_W=\frac{\pi\alpha}{\sqrt{2}G_F'm_Z^2}(1-\delta)~,
\label{s2W2}
\end{equation}
and thus $\sin^2\theta_W=0.23107\mp 0.00005$, and $\shat^2_Z=0.23130\mp 0.00007$.
The value is slightly smaller than that of case 1, so a precise measurement of $\sin^2\theta_W$
would determine which scenario is realized in our universe.
One can also construct far more complicated cases with various ultimate speeds.
\par
In conclusion, the "$c'$-scenario" presented in this work can explain the superluminal
OPERA neutrinos.
It happens that $c$ and $c'$ are quite degenerate so it is possible that the effect of $c'$ 
might have not been detected up to now.
It was shown that new effects slightly shift the weak mixing angle but still within the errors.
Much more precise determination of the electroweak observables would confirm or refute the scenario.

\end{document}